\documentclass[superscriptaddress,amsmath,amsfonts,preprint]{revtex4-1}
\usepackage{graphicx}
\usepackage[normalem]{ulem}
\usepackage[colorlinks=true]{hyperref}
\begin{document}

\title{Transport Phase Diagram and Anderson Localization in Hyperuniform Disordered Photonic Materials}

\author{Luis S.\ Froufe-P\'erez}
\affiliation{Department of Physics, University of Fribourg, CH-1700 Fribourg, Switzerland}

\author{Michael Engel}
\affiliation{Institute for Multiscale Simulation, Friedrich-Alexander-University Erlangen-Nuremberg, 91052 Erlangen, Germany}

\author{Juan Jos\'e S\'{a}enz}
\affiliation{Donostia International Physics Center (DIPC), Paseo Manuel Lardizabal 4, 20018 Donostia-San Sebasti\'an, Spain}
\affiliation{IKERBASQUE, Basque Foundation for Science, 48013 Bilbao, Spain}

\author{Frank Scheffold}
\affiliation{Department of Physics, University of Fribourg, CH-1700 Fribourg, Switzerland}

\date{\today}

\begin{abstract}{\bf Hyperuniform disordered photonic materials (HDPM) are spatially correlated dielectric structures with unconventional optical properties \cite{Torquato_J_App_Phys_2008,Torquato2015}. They can be transparent to long-wavelength radiation while at the same time have isotropic band gaps in another frequency range \cite{Florescu_PNAS_2009,Leseur2016}. This phenomenon raises fundamental questions concerning photon transport through disordered media. While optical transparency is robust against recurrent multiple scattering, little is known about other transport regimes like diffusive multiple scattering or Anderson localization \cite{chabanov2000statistical}. Here we investigate band gaps, and we report Anderson localization in two-dimensional stealthy HDPM using numerical simulations of the density of states and optical transport statistics. To establish a unified view, we propose a transport phase diagram. Our results show that, depending only on the degree of correlation, a dielectric material can transition from localization behavior to a bandgap crossing an intermediate regime dominated by tunneling between weakly coupled states.}
\end{abstract}
\maketitle
Light propagation through a dielectric medium is determined by the spatial distribution of the material. Photons scatter at local variations of the refractive index. For a periodically organized system, interference dominates light transport and is responsible for optical phenomena in opal gems and photonic crystals~\cite{Joannopoulos_book}. In random media, transport becomes diffusive through successive scattering events. The characteristic length scale over which isotropic diffusion takes place is the transport mean free path. Material thicker than the mean free path appears cloudy or white. However, when scattering centers are locally correlated, diffraction effects can be significant. The description of light transport then becomes a challenging problem with many applications, such as the transparency of the cornea to visible light~\cite{benedek1971theory},
the strong wavelength dependence of the optical thickness of colloidal suspensions~\cite{rojas2004photonic} and amorphous photonic structures~\cite{reufer2007transport,garcia2007photonic}, and structural colors in biology~\cite{prum1998coherent}. 
Critical opalescence
and the relatively large electrical conductivity of disordered liquid metals~\cite{ashcroft1966structure} are closely related phenomena.

In the weak scattering limit photon transport is diffusive and can be described by a local collective scattering approximation, which states that the mean free path $l_t =(\rho \sigma_t)^{-1}$ is inversely proportional to the number density of scatterers $\rho$ and to the effective transport cross section~\cite{ashcroft1966structure,fraden1990multiple,conley2014light} 
\begin{equation}
 \sigma_t = \int \frac{d \sigma}{d \Omega} S(k_\vartheta) (1-\cos \vartheta) d \Omega. \label{mfp}
\end{equation}
Here $d\sigma/d\Omega$ is the differential cross section for an isolated scatterer and $k_\vartheta = 2k \sin(\vartheta/2)$ is the momentum transfer. This equation relates positional correlations of the optical medium to transport via the structure factor $S(k_\vartheta)$. In the past, the local collective scattering approximation has been applied to dense and strongly scattering media~\cite{conley2014light}, however the validity of this approach is limited as it does not include near-field corrections and recurrent scattering~\cite{wiersma1995experimental,naraghi2015near}. Clearly, the recently discovered isotropic bandgaps~\cite{Florescu_PNAS_2009,Froufe_PRL_2016} in HDPM cannot be derived from Eq.~\ref{mfp}.

For uncorrelated or fully random media with $S(k_\vartheta)\equiv 1$ optical transport properties are well understood. 
According to the single parameter scaling (SPS) hypothesis one expects a transition from diffuse scattering to Anderson localization~\cite{abrahams1979scaling,markovs2006numerical}. Statistical properties of transport are governed by a single parameter that can be expressed as the ratio of a characteristic (localization) length $\xi$ to system size $L$. While for $L/\xi \ll 1$ transport is diffusive, SPS predicts a crossover to the Anderson localization regime $L/\xi \gg 1$ for both disordered wires (quasi-one-dimensional system) and two-dimensional systems with any amount of disorder. 

For correlated disordered media, over the last years, attention focused on the emergence of optical transparency~\cite{Torquato_J_App_Phys_2008,Leseur2016} and on photonic bandgaps (PBGs) in two-dimensional high-refractive-index disordered materials~\cite{Florescu_PNAS_2009,Froufe_PRL_2016,Edagawa2008}. In particular, the concept of stealthy hyperuniformity as a measure for the hidden order in amorphous materials has drawn significant attention~\cite{Torquato_J_App_Phys_2008,Torquato2015}.
HDPM are disordered, but uniform without specific defects, and the density of states can be strictly zero. A key parameter controlling structural correlations of HDPM, and thus the band gap width, is the stealthiness parameter $\chi$, defined as the ratio between the number of constrained degrees of freedom to the total number of degrees of freedom~\cite{Torquato_J_App_Phys_2008}. 
It is currently not known what effect $\chi$ has on the statistics of wave transport in HDPM outside the gap and whether the transport properties in HDPM can be understood in terms of SPS. Neither is anything known about the transport properties for frequencies $\nu$ near the band edge, a regime where disordered crystalline materials exhibit a rich and complex transition towards Anderson localization~\cite{john1996localization}. The main goal of our work is to address these fundamental questions and to propose a $\chi$--$\nu$ phase diagram for wave transport through ideal hyperuniform disordered dielectric materials in two dimensions. 

\begin{figure}
\includegraphics[width=0.8\columnwidth]{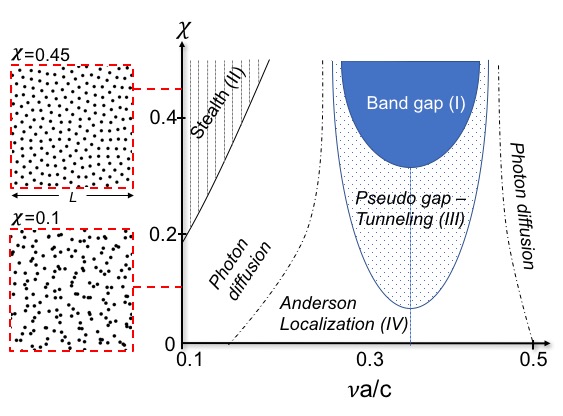}
\caption{\label{fig:phasediagram} Photonic transport phase diagram for hyperuniform disordered dielectric materials. A box of size $L\times L$ is filled with high refractive index ($\varepsilon=11.6$) cylinders, radius $0.189 a$, at a filling fraction of $11.2\%$~\cite{Froufe_PRL_2016}. In the central region (I) a wide photonic gap opens inhibiting the propagation of TM polarized electromagnetic waves~\cite{Florescu_PNAS_2009,Froufe_PRL_2016}. In this region the density of states is exactly zero and evanescent light waves decay exponentially over distances shorter than the characteristic structural length scale $\xi<a$. In the stealth region (II), on the left side of the band gap, the material is transparent. In the vicinity of the gap, for smaller values of $\chi$, the density of states is suppressed but non-zero (III), and evanescent waves can tunnel between isolated states with a decay length $\xi\sim a$. Sufficiently far from the gap diffuse light transport may cross over to strong Anderson localization. The transition from diffusive to localizated transport (dotted line) is system-size-dependent and we expect Anderson localization (IV) in the limit of sample sizes $L > \xi > a$. The degree of hyperuniformity is denoted $\chi$. $\nu$ is the optical frequency, $c$ the vacuum speed of light, and $a$ is the mean distance between scatterers. Eventually, at high $\chi$ (not shown), we enter the crystal regime~\cite{Florescu_PNAS_2009,Froufe_PRL_2016}.} 
\end{figure}

A relatively simple example of HDPM is a collection of infinite parallel cylinders of high-refractive index that are distributed according to a two-dimensional stealthy hyperuniform point pattern (SHU, see the insets in Fig.~\ref{fig:phasediagram}). The surface area $A$ and the number of points $N$ define a characteristic length $a$ through the number density $\rho=N/A\equiv a^{-2}$. When the electric field is parallel to the cylinder axes (TM polarization), the propagation of light with a wave vector perpendicular to the axes can be described by a two-dimensional Schr\"odinger-like scalar equation with identical scatterers. This setup provides an ideal laboratory to explore the statistical properties of wave transport in two-dimensional HDPM. The normalized density of states (NDOS) as a function of frequency for a two-dimensional SHU pattern of monodisperse high refractive index cylinders has been reported for different values of $\chi$~\cite{Florescu_PNAS_2009,Froufe_PRL_2016}.
 
Based on the NDOS and the results for two-dimensional disordered crystals~\cite{deych2001single,prior2005conductance}, we first make a hypothesis about the transport phase diagram of two-dimensional HDPM in Fig.~\ref{fig:phasediagram}. For strongly correlated, nearly crystalline materials (high $\chi$), stealth (transparent) and gap intervals are adjacent to each other, reminiscent of the conduction band and stop gap for electrons and photons in crystals. Lowering the degree of hyperuniformity to $\chi < 0.5$, the gap and stealth intervals shrink. In the newly accessible intervals in-between the stealth and the gap regions, the material possesses a reduced density of states and displays significant scattering.
We expect a tunneling-regime (non-SPS) for frequencies near the full band gap and a diffusive scattering regime further apart. The tunneling regime is intimately related to the pseudo gap in which the density of states is low, but finite. Once the density of states is sufficiently high, far away from the gap, diffusive transport sets in and a crossover to classical Anderson localization should occur as $L>\xi$.
For high frequencies we also expect diffusive transport, independent of $\chi$, since in this spectral region the scattering strength decays below the threshold where a gap opens.

\begin{figure}
\includegraphics[width=0.7\columnwidth]{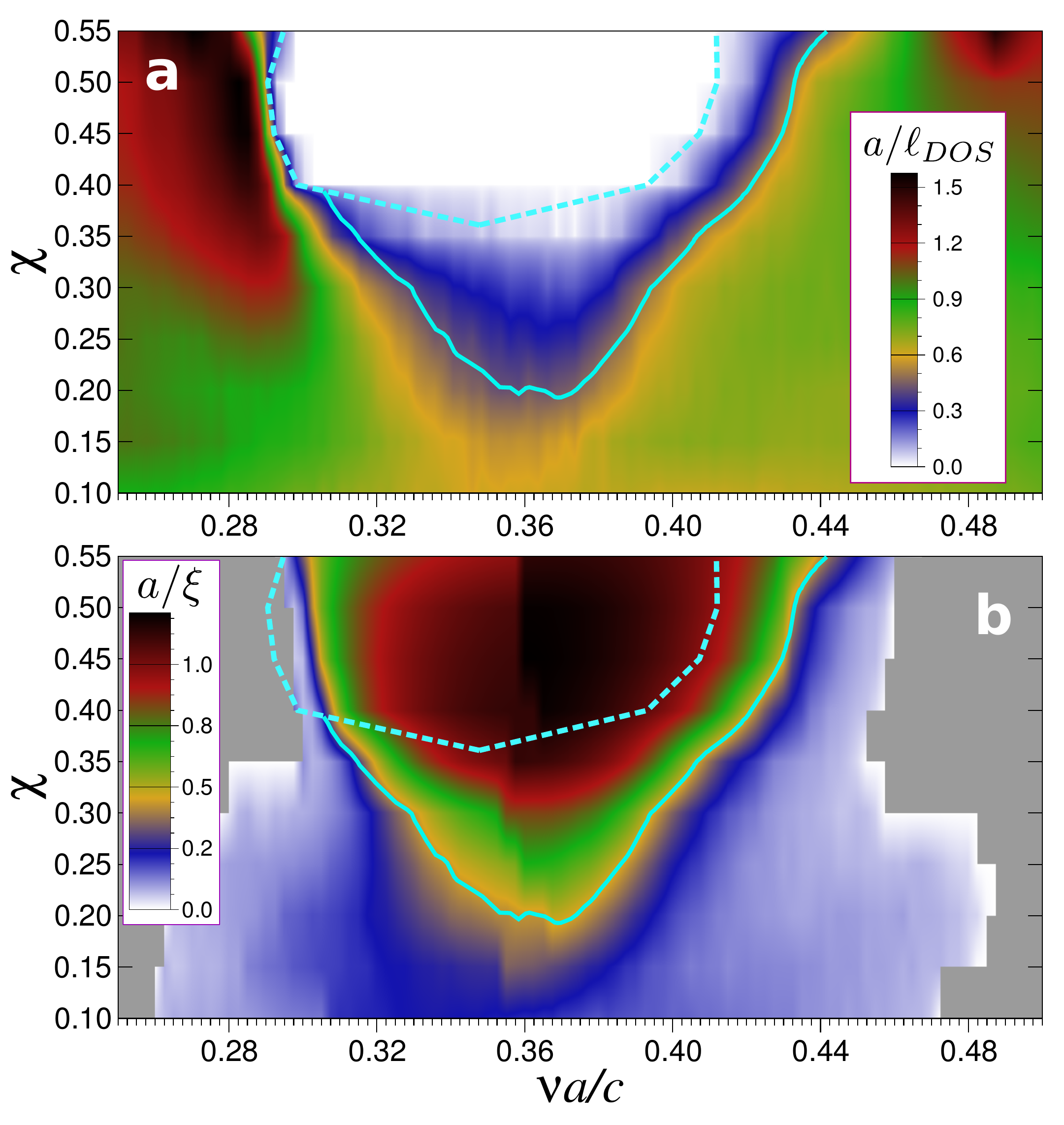}
\caption{\label{fig:figDOS} Ratio of the characteristic size $a$ of the system to (a)~the mean distance between states $\ell_{\text{DOS}}$ and (b)~the decay length $\xi$ (color maps) as a function of frequency and stealthiness for SHU systems. The decay length can only be extracted if $\xi<L_\text{max}$. Gray areas in (b) thus correspond to systems where the condition $\langle \ln(g)\rangle \le\ln(1/2)$ has not been reached. The dashed line delineates the full band gap and the solid line indicates $\xi= \ell_{\text{DOS}}$ in Figs. \ref{fig:figDOS}-\ref{fig:DMPK}.}
\end{figure}

Although the calculation of the standard transport mean free path can be useful to describe transport processes in disordered correlated media in frequency ranges where the system is almost transparent or transport is diffusive~\cite{conley2014light,Leseur2016}, it loses its meaning in a region where the density of states is zero. Identifying the different transport regimes requires the numerical solution of the full multiple scattering problem in a wide spectral range. 
We calculate the decay of the intensity of the wave fields along the propagation direction and the full statistics of wave transport. First, we generate a statistical ensemble of 1000 spatial patterns of monodisperse high refractive index cylinders for each value of $\chi$ as described in~\cite{Froufe_PRL_2016}.
Our spatial two-dimensional point patterns show stealthy hyperuniformity, which means that not only long-range density fluctuations are suppressed,
$S(k_\vartheta \rightarrow0)=0$, but the structure factor vanishes over a finite range, $S(k_\vartheta<k_{c})=0$. The critical wave number $k_c$ sets the phase boundary of the stealth phase~\cite{Leseur2016,Froufe_PRL_2016}. 
With these parameters a PBG opens at a central frequency $\nu_{0}a/c\simeq0.35$, and the PBG expands to a maximum width of $\Delta\nu/\nu_{0}\simeq45\%$~\cite{Froufe_PRL_2016}. Following prior work on disordered crystals~\cite{deych2001single,prior2005conductance} we introduce the length
\begin{equation}
\ell_{\text{DOS}}(\nu,\chi)=\frac{a}{\sqrt{\text{NDOS}(\nu,\chi)}}
\end{equation}
as an effective measure of the mean distance between states. Fig.~\ref{fig:figDOS}(a) shows a map of $a/\ell_{\text{DOS}}$. We can clearly identify the gap with no states and infinite $\ell_{\text{DOS}}$ (white region bounded by dashed line) and a pseudo gap region where the NDOS is significantly reduced, $\ell_{\text{DOS}}\gg a$.

\begin{figure}
\includegraphics[width=.75\columnwidth]{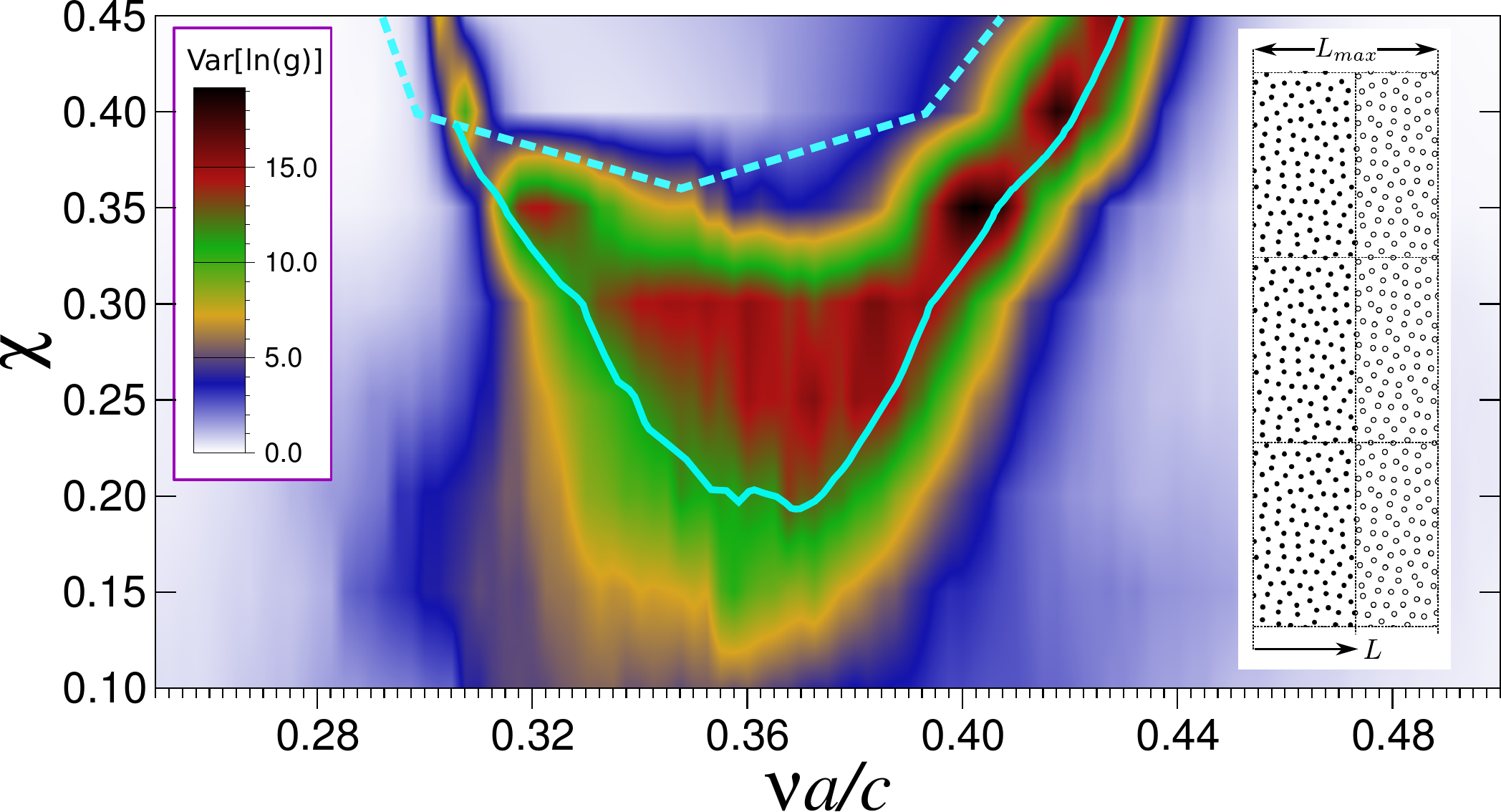}
\caption{\label{fig:log_g_flucts} Fluctuations of the logarithm of the conductance $\text{Var}(\ln(g))$ after TM-polarized electromagnetic waves passed through the system, $L=L_\text{max}$. The inset shows the simulation setup of the SHU pattern with dielectric cylinders using periodic boundary conditions in transverse directions.}
\end{figure}

To obtain the characteristic decay length $\xi$ and to sample the statistical properties of transport, we apply the well-known generalized scattering matrix (GSM) method~\cite{a2004improved} to a system of fixed width $L_\text{max}$ as a function of the thickness $L<L_\text{max}$ in propagation direction. Periodic boundary conditions (inset in Fig.~\ref{fig:log_g_flucts}) define a set of transversal propagation channels. The optical analog $g$ of electrical conductance can be computed from Landauer's formula. If $T_{ij}$ is the intensity transmitted from incoming channel $j$ to outgoing channel $i$, then
the conductance of the system is $g=\sum_{ij}T_{ij}$. The exponential decay length $\xi$ of the conductance is determined using a fit to $\langle \ln(g(L))\rangle = -2L/\xi$~\cite{Beenakker_RMP_1997}.
Fluctuations are suppressed by performing an ensemble average $\langle\cdot\rangle$ over 1000 realizations for each frequency and stealthiness. Fig.~\ref{fig:figDOS}(b) shows a map of $a/\xi$.
The solid line corresponds to the boundary $\ell_{\text{DOS}}=\xi$. Based on studies of two-dimensional disordered crystals~\cite{prior2005conductance}, we expect this line to indicate the boundary between non-SPS and SPS regions.

So far we identified regions with exponential decay of the conductance but did not reveal
the physical mechanism responsible for this rapid decay. To this end we analyze the optical transport statistics~\cite{chabanov2000statistical}.  Fig.~\ref{fig:log_g_flucts} shows a color map of fluctuations in the logarithmic conductance $\text{Var}(\ln(g))$ for transmission through the whole system. As expected, regions associated with a PBG do not show substantial fluctuations despite a small decay length $\xi$ signaling transport by direct tunneling through the whole sample. Outside the gap, large fluctuations are most pronounced in the pseudo gap regime, $0< \xi/l_{\text{DOS}} \lesssim 1$ indicating tunneling-like transport~\cite{prior2005conductance}.
Eventually, further away from the gap, in the regime with $\xi/l_{\text{DOS}}>1$, fluctuations decay rapidly.

\begin{figure}
\includegraphics[width=.64\columnwidth]{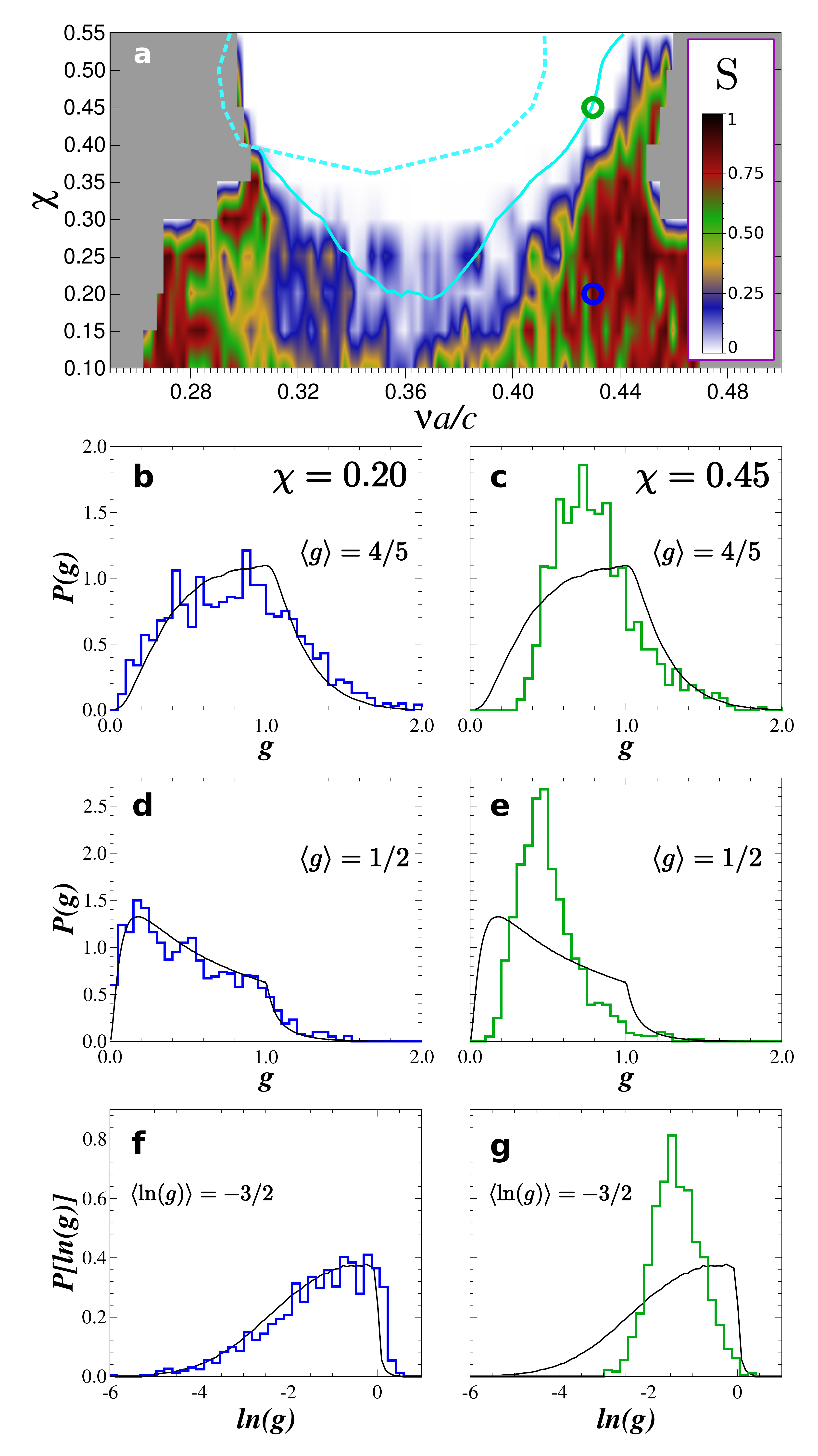}
\caption{\label{fig:DMPK} (a)~Similarity map S$(\nu,\chi)$ for the conductance distribution at $g=1/2$. High values indicate that the conductance distribution $P(g=1/2)$ is similar to the one predicted by DMPK. The function S$(\nu,\chi)$ is bounded in $[0,1]$. (b-g)~Conductance distribution for two typical situations, in the SPS regime, $\chi=0.20$, and in the tunneling regime inside the pseudo gap, $\chi=0.45$. Numerical results are compared to predictions of DMPK (black lines).}
\end{figure}

To discriminate between tunneling-like transport and Anderson localization, we consider the statistical distribution $P(g)$ of the conductance. A quantitative description for transport fluctuations in quasi-one-dimensional systems is given by the Dorokhov-Mello-Pereyra-Kumar (DMPK) equation~\cite{dorokhov1982transmission,mello1988macroscopic,froufe2002conductance}.
DMPK predicts a crossover between the diffusive regime and the localized regime. While $P(g)$ is Gaussian in the diffusive regime, it has a peculiar shape at the onset of the localized regime corresponding to $\langle g \rangle \approx 1/2$ with a marked discontinuity in the first derivative of the distribution and a sharp cut-off beyond $g=1$~\cite{froufe2002conductance,muttalib1999one}. As $\langle g \rangle$ decreases, this cut-off eventually leads to one-sided log-normal distributions for the conductance. 
Although the DMPK results were derived for the quasi-one-dimensional case, the characteristic shape of $P(g)$ at the onset of the localized regime
was exactly reproduced by numerical simulations in two-dimensional disordered systems in the SPS regime~\cite{somoza2009crossover}. 
We apply a similarity analysis of the numerical $P(g)$ results in our HDPM system and the DMPK result~\cite{froufe2002conductance} to identify regions
where Anderson localization occurs. A map of the similarity S$(\nu,\chi)$ is shown in Fig.~\ref{fig:DMPK}.
Outside the gap and pseudo gap regions $S>0.5$ and light transport evolves from diffusion, where the conductance distributions are Gaussian, to Anderson localization, where the distribution is one-side log-normal.
The behaviour of $P(g)$ in the pseudo-gap region indicates remarkable differences in the statistical properties, which suggests that SPS is violated in this region.


In conclusion, our numerical results fully support the proposed transport phase diagram in Fig.~\ref{fig:phasediagram}.
We expect that our findings are not restricted to SHU correlations, encoded by the parameter $\chi$, but equally apply to other types of uniform spatial correlations~\cite{Froufe_PRL_2016}. A similar transport phase diagram can be established in three dimensions.
In this case a sharp phase boundary, known as the mobility edge and set by the Ioffe-Regel criterion $k l_t \lesssim 1$~\cite{john1996localization}, is expected between photon diffusion and Anderson localization (dash-dotted line in Fig.~\ref{fig:phasediagram}).
An important direction for future work are the implications of the phase diagram for electronic transport. It has been argued that SHU plays a role in the formation of electronic band gaps, for example in amorphous silicon~\cite{xie2013hyperuniformity}. However,  the influence of structural correlations and hyperuniformity on electron transport and localization in two or higher dimensions is far from being understood~\cite{izrailev2012anomalous}.  

\section{Methods} 
\noindent\textbf{\textit{Generation of stealthy hyperuniform point pattern.}}
We employ a simulated annealing relaxation scheme to generate disordered SHU patterns with $S(k)<10^{-6}$ for $k<k_c(\chi)$ as described in our previous work, ref.  ~\cite{Froufe_PRL_2016}. Patterns below the critical parameter $\chi\sim 0.55$, above which quasi-long-range order gradually appears~\cite{Torquato2015}, already show significant short-range order. The point patterns are decorated with dielectric cylinders as described in the text.
\newline

\noindent\textbf{\textit{Band structure calculation.}}
We calculate the normalized photonic density of states (NDOS) using the supercell method~\cite{Joannopoulos_book} implemented in the open source code MIT Photonic Bands~\cite{Johnson2001_mpb}. The supercell is repeated periodically and the band structure calculated by following the path $\Gamma\rightarrow M\rightarrow X\rightarrow\Gamma$ in reciprocal space. The procedure is described in detail in our reference~\cite{Froufe_PRL_2016}. Nearly identical strategies have also been applied by others~\cite{Florescu_PNAS_2009}. 
\newline

\noindent\textbf{\textit{Generalized scattering matrix (GSM) method.}}
We use the improved generalized scattering matrix method. A summary of the method is provided in ~\cite{a2004improved}. In the GSM the systems is discretized in slices in the propagation direction. For each slice the wave equation is solved and the scattering matrix calculated. By sequentially combining the corresponding scattering matrices, the total scattering matrix of the system up to a length $L$ is obtained.
\newline

\noindent\textbf{\textit{Dorokhov-Mello-Pereyra-Kumar (DMPK) equation.}}
The Dorokhov-Mello-Pereyra-Kumar (DMPK) equation is a Fokker-Plank equation describing the evolution of coherent wave transport statistics as a function of the ratio of the system length to the transport mean free path. It was shown to describe quantitatively the transport properties of disordered quasi-one-dimensional systems. There is evidence that the DMPK distributions retain the main properties of the conductance distributions in the metallic, critical and localized regime also in higher dimensions~\cite{muttalib1999one,markovs2006numerical}. The DMPK distributions were obtained as described in ref.~\cite{froufe2007statistical}, see also supplementary material.
\newline

\noindent\textbf{\textit{Similarity analysis.}}
The similarity between the conductance distributions of correlated systems and the ideal DMPK distribution is characterized by a similarity function. We first create numerically a finite sample of DMPK conductance histograms with a certain bin size. We quantify these differences between the ideal DMPK sample and the results obtained by the GSM by calculating a squared distance distribution function $P(D^{2})$. The squared distance between two distributions $P_{i}(g)$ and $P_{j}(g)$ is given by
\begin{equation}
D_{i,j}^{2}\equiv\int dg\,[P_{i}(g)-P_{j}(g)]^{2}.
\end{equation}
In other contexts the squared distance $D^{2}$ is denoted $\chi^{2}$ or chi-squared. We do not make use of this notation to avoid confusion with the stealthiness parameter $\chi$. For a set of different, but otherwise statistically equivalent, finite-sampling distributions $P_{i}(g)$ we can define the distribution of squared distances $P(D^{2})$ considering all distances $D_{i,j}$ corresponding to all pairs with $i\ne j$. To assess the similarity between $P(D^{2})$ and the reference $P_\text{DMPK}(D^{2})$ distribution, we compute the area below both distributions. Similarity is defined as
\begin{equation}
S\equiv\int_{0}^{\infty}dx\,\min\{ P_\text{diff}(x),P_\text{DMPK}(x)\} \textrm{.}\label{eq:320}
\end{equation}
By this definition $S$ is bounded in $[0,1]$ (see also supplementary material).


\section*{Acknowledgements} This research was supported by the Swiss National Science Foundation through the National Center of Competence in Research Bio-Inspired Materials and through Project No.\ 149867 and 169074 (L.F.\ and F.S.), Spanish Ministerio de Econom\'ia y Competitividad (MICINN) and European Regional Development Fund (ERDF) through Project FIS2015-69295-C3-3-P and the Basque Dep.\ de Educaci\'on through Project PI-2016-1-0041 (J.J.S.). M.E.\ acknowledges funding by Deutsche Forschungsgemeinschaft through the Cluster of Excellence Engineering of Advanced Materials and support from the Central Institute for Scientific Computing (ZISC) and the Interdisciplinary Center for Functional Particle Systems (IZ-FPS) at FAU Erlangen-Nuremberg.

\section*{References}
\bibliographystyle{naturemag}





\clearpage

\clearpage

\clearpage

\clearpage

\clearpage

\end{document}